\documentclass[amsmath,10pt,showpacs,twocolumn,superscriptaddress,aps,prl,floatfix]{revtex4-1}

\usepackage{graphicx}
\usepackage{multirow}
\usepackage[usenames]{color}
\usepackage{nicefrac}
\usepackage{amsmath}
\usepackage{float}
\usepackage{silence}
\begin{document}

\title{Superfluid transport dynamics in a capacitive atomtronic circuit}
\author{Aijun Li}
\affiliation{Department of Physics, Georgia Southern University,
Statesboro, GA 30460--8031 USA}
\affiliation{Key Laboratory of Coherent Light, Atomic and Molecular Spectroscopy, 
College of Physics, Jilin University}
\author{Stephen Eckel}
\affiliation{Joint Quantum Institute, National Institute of Standards 
and Technology and the University of Maryland, Gaithersburg, MD 20899, USA}
\author{Benjamin Eller} 
\affiliation{Department of Physics, Georgia Southern University,
Statesboro, GA 30460--8031 USA}
\author{Kayla E.\ Warren} 
\affiliation{Department of Physics, Georgia Southern University,
Statesboro, GA 30460--8031 USA}
\author{Charles W.\ Clark}
\affiliation{Joint Quantum Institute, National Institute of Standards 
and Technology and the University of Maryland, Gaithersburg, MD 20899, USA}
\author{Mark Edwards} 
\affiliation{Department of Physics, Georgia Southern University,
Statesboro, GA 30460--8031 USA}
\affiliation{Joint Quantum Institute, National Institute of Standards 
and Technology and the University of Maryland, Gaithersburg, MD 20899, USA}
\date{\today}

\begin{abstract}
We simulate transport in an atomtronic circuit of a Bose-Einstein condensate that flows from a source region into a drain through a gate channel. The time-dependent Gross--Pitaevskii equation (GPE) solution matches the data of a recent experiment.  The atomtronic circuit is found to be similar to a variable--resistance RLC circuit, which is critically damped at early times and shows LC oscillations later. The GPE also predicts atom loss from the drain. Studies of the dependence of condensate transport upon gate parameters suggest the utility of the GPE for investigation of atomtronic circuits.
\end{abstract}


\maketitle


The ``lumped abstraction'' model provides an interface between the physics of electromagnetism and the engineering of electronic circuits. By attributing ideal independent macroscopic properties (e.g. resistance, inductance, and capacitance) to the individual components a circuit, the lumped abstraction model makes possible the design of highly complex functional  circuits.~\cite{Foundations_Agarwal_Lang,art_of_electronics}.  A major challenge in the emerging field of atomtronics is the establishment of a comparable interface for the design of atom circuits. Lee {\em et al.}\,\cite{SciRep.3.1034} have found equivalents of electronic resistance, capacitance and inductance in a simple atomtronic circuit. We believe that the time--dependent Gross--Pitaevskii equation can be useful in both determining the validity of an atomtronic lumped abstraction model and, if valid, determining the values of circuit parameters. Here we test this hypothesis by applying it to a recent atomtronic experiment~\cite{nist_paper}.

A typical example of an atom circuit consists of a Bose--Einstein condensate (BEC) harmonically confined to a horizontal plane by a red--detuned laser and arbitrarily confined within the plane by a combination of red-- and blue--detuned lasers ~\cite{Henderson, Gaunt2012,Pasienski,Ryu2015,2016arXiv160504928G}.  Atomtronic devices analogous to batteries, diodes, transistors, and fundamental logic gates have been proposed~\cite{PhysRevA.70.061604, PhysRevLett.103.140405, PhysRevA.75.023615,Caliga2016,Caliga2016a,Zhang2015,Ott2015}.  Atom circuits in the form of a BEC confined in a ring geometry have been studied as potential rotation sensors~\cite{PhysRevLett.106.130401, PhysRevLett.110.025302,PhysRevLett.113.045305,Eckel2014,1367-2630-17-4-045023,Bell2016,Kumar2016,Wang2015,Mathey2016}. 

Recently, a series of experiments was conducted in which a gas of thermal atoms~\cite{SciRep.3.1034} and Bose--Einstein--condensed atoms~\cite{nist_paper} were confined in a quasi--2D potential consisting of two wells connected by a channel.  The atoms were initially confined the {\em source} well and then released to flow down the channel into the {\em drain} well.  The difference between the number of atoms in the source, $N_{S}(t)$, and the number in the drain, $N_{D}(t)$, normalized by their sum, the number imbalance $\Delta N(t)=(N_{S}-N_{D})/(N_{S}+N_{D})$, as a function of time was inferred from image data.  Similar experiments in Fermi gases have been recently reported~\cite{Husmann1498,Nature.517.64}.

\begin{figure}[htb]
\begin{center}
\includegraphics[width=4in]{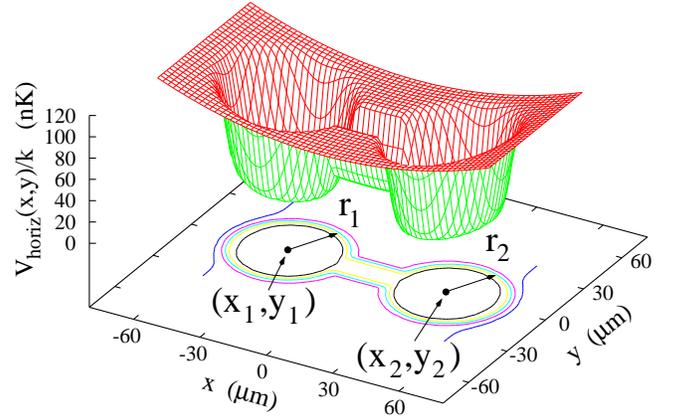}
\end{center}
\caption{(color online) The potential in the horizontal ($xy$) plane consists of the mask potential and the $xy$ part of the (harmonic) sheet potential.  The mask potential is zero inside hard--walled circular wells (with centers $(x_{k},y_{k})$ and radii $r_{k}$ where $k=1,2$) and is equal to $V_{d}$ outside; inside the channel the mask potential is harmonic along the $y$ direction plus a constant step.}
\label{horizontal_potential}
\end{figure} 
In this paper we present simulations of the experiment of Ref.\,\cite{nist_paper} using the time--dependent Gross--Pitaevskii (GP) equation and show that it quantitatively captures the physics of the evolution of the number imbalance. We then describe the features of the GP--predicted transport dynamics in terms of a variable--resistance RLC circuit.  The dynamics includes atom loss from the drain and we provide a possible mechanism for this loss.  Finally we summarize our study of the transport dynamics dependence on the length and effective width of the channel.

The time--dependent GP equation~\cite{gp_gross,gp_pitaevskii} has the form
\begin{equation}
i\hbar\frac{\partial\Psi}{\partial t} = 
\left[
-\frac{\hbar^{2}}{2M}\nabla^{2} +
V_{\rm trap}({\bf r},t) + 
gN\left|\Psi\right|^{2}
\right]\Psi({\bf r},t),
\label{gpe}
\end{equation}
where $M$ is the mass of a condensate atom, $g=4\pi\hbar^{2}a_{s}/M$ measures the strength of binary atom scattering where $a_{s}$ is the $s$--wave scattering length, $N$ is the number of condensate atoms, and $V_{\rm trap}({\bf r},t)$ is the trap potential in which the condensate atoms move.  In simulating the experiment in Ref.~\cite{nist_paper} we found that obtaining agreement with the data strongly depended on careful modeling of the trap potential.

The optical dipole trap present in Ref.~\cite{nist_paper} was produced by the superposition of a horizontal, planar red--detuned light sheet and a vertical, blue--detuned, Gaussian laser beam (the ``mask beam) partially blocked by a dumbbell--shaped mask.  We modeled this as a 3D harmonic ``sheet'' potential plus a 2D dumbbell--shaped ''mask'' potential plus a ``gate'' potential, a high step function that blocks the channel only during condensate formation. The mask potential was, in turn, modeled as the superposition of ``well'' and ``channel'' potentials.  The full model potential thus had the form
\begin{eqnarray}
V_{\rm trap}({\bf r},t) 
&=& 
\frac{1}{2}M
\left(
\omega_{\rm sh,x}^{2}x^{2} + 
\omega_{\rm sh,y}^{2}y^{2} +
\omega_{\rm sh,z}^{2}z^{2}
\right)\nonumber\\8
&+&
V_{\rm well}(x,y) + 
V_{\rm channel}(x,y)\nonumber\\ 
&+&
\epsilon(t) V_{\rm gate}(x).
\label{trap_pot}
\end{eqnarray}
For the gate potential, $\epsilon=1$ during condensate formation and is zero otherwise. The $z$ axis is vertical, the $x$ axis lies along the line joining the two well centers, and the $y$ axis is perpendicular to the channel (see Fig.~\ref{horizontal_potential}).

\begin{figure}
\begin{center}
\includegraphics[width=3.25in]{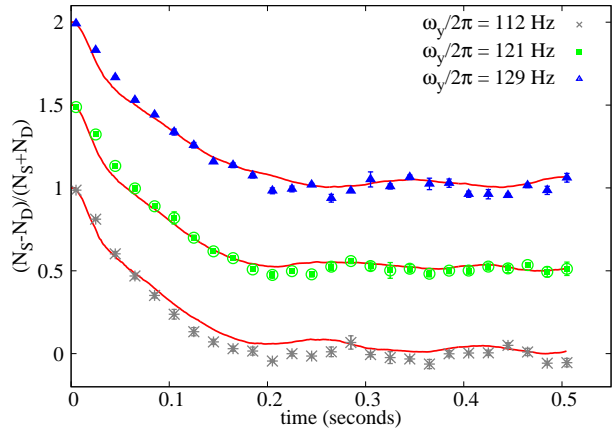}
\end{center}
\caption{(color online) Atom number imbalance (with error bars) between source and drain wells versus time from Ref.~\cite{nist_paper} for different channel transverse trapping frequencies, $\omega_{y}/2\pi$: 112 Hz (gray x's, bottom curve); 121 Hz (green circles, middle curve); and 129 Hz (blue triangles, top curve). For all cases the channel length was $L_{c}=26\ \mu{\rm m}$ and the well depths where $V_{0}/k=83$ nK where $k$ is the Boltzmann constant.  The red solid curves show the GP simulation.  The number of condensate atoms was taken to be $N=$ 480,000. The middle and top curves have been vertically offset by 0.5 and 1.0, respectively, for clarity.}
\label{gpe_exp_compare_fig}
\end{figure} 

The well and channel potentials had the form
\begin{eqnarray}
V_{\rm well}(x,y)
&=& 
V_{d}\sum_{k=1,2}\frac{1}{2}
\left[1+\tanh\left(\frac{\rho_{k}(x,y)-r_{k}}{b}\right)\right],\nonumber\\
\rho_{k}(x,y)
&\equiv&
\sqrt{(x-x_{k})^{2}+(y-y_{k})^{2}},
\quad
k = 1,2,\nonumber\\
V_{\rm channel}(x,y) 
&=& 
V_{\rm step}+
\frac{1}{2}M\omega_{y}^{2}y^{2}.
\end{eqnarray}
The well potential was zero inside two wells with centers at $(x_{k},y_{k})$ and radii $r_{k}$ ($k=1,2$) and equal to the well depth, $V_{d}$, outside.  The hardness of the well edges was controlled by the value of $b$.  The channel potential was modeled as a step plus a harmonic oscillator along $y$.  The channel length was defined by $L_{c}=(x_{2}-x_{1})-(r_{2}+r_{1})$ and the mask potential was set equal to $\min(V_{\rm well},V_{\rm channel})$ when $x_{1} \le x \le x_{2}$ and equal to $V_{\rm well}$ otherwise. 

The sheet--potential frequencies were determined to be $\omega_{\rm sh,x}/2\pi=10$ Hz, $\omega_{\rm sh,y}/2\pi\approx 0$ Hz, and $\omega_{\rm sh,z}/2\pi=529$ Hz. The well radii were $r_{1}=r_{2}= 24\ \mu$m, the well depth was $V_{d}/k=83$ nK, and the hardness parameter was $b=0.2\ \mu$m.  The well centers were separated by $x_{2}-x_{1}=74\ \mu$m. In the channel $V_{\rm step}/k\approx 20$ nK and $\omega_{y}/2\pi$ varied between 110 and 130 Hz (see Fig.~\ref{gpe_exp_compare_fig}). More details are given in the Supplementary Materials.  

The solution of the 3D time--dependent GP equation was approximated using the hybrid Lagrangian Variational Method (HLVM)~\cite{PhysRevE.86.056710}. This approach is valid when there is strong confinement to a planar region. The solution is assumed to be a product of an unrestricted function in the plane and a Gaussian in the strongly confined direction. The HLVM equations were solved using the split--step, Crank--Nicolson algorithm~\cite{adhikari} in a $150\,\mu{\rm m}\times 75\,\mu{\rm m}$ box with a space step of $\Delta x = \Delta y= 0.09375\ \mu$m.  The initial, variationally stable condensate was determined as described in Ref.~\cite{PhysRevE.86.056710} and evolved for a total of 0.5 seconds using 700,000 time steps.  Condensate transport dynamics was modeled by integrating the GP equation on the same with $\epsilon=0$ in Eq.\,(\ref{trap_pot}).

\begin{figure*}
\begin{center}
\includegraphics[width=6.75in]{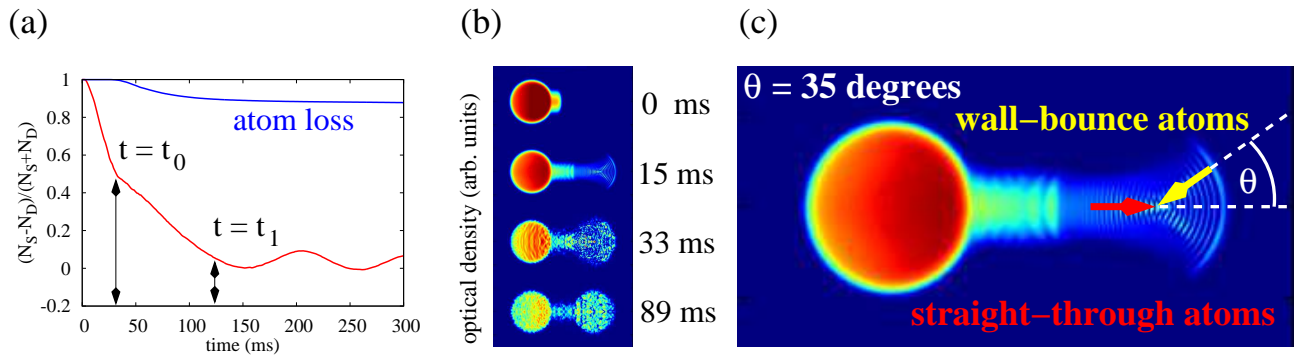}
\end{center}
\caption{(color online) (a) Number imbalance, $\Delta N(t)$, for a dumbbell potential with channel length $L_{c}=20\,\mu\mathrm{m}$, Thomas--Fermi width of $w_{\mathrm{TF}}\approx 22\,\mu\mathrm{m}$ (or transverse frequency $\omega_{y}/2\pi=63$ Hz), well depth of $V_{d}/k=97$ nK, and $N=436\mathrm{,}000$ atoms. The bottom (red) curve is the GP solution. The top (blue) curve shows the fraction of initial number of atoms remaining in the dumbbell area versus time. (b) Dumbbell optical density snapshots at four times during filling of the drain well. (c) Enlargement of the dumbbell optical density at $t=15$ ms.  Atoms that flow straight along the dumbbell axis can collide with atoms that bounce off the drain wall at angle $\theta$.}
\label{gpe_deltaN}
\end{figure*} 

As described below, the evolution of the system exhibited atom loss from the dumbbell area.  This significantly degraded the GP simulation because atoms leaving the dumbbell region were reflected from the grid boundary back into the dumbbell.  Thus, at each time step, the GP solution was multiplied by a windowing function having unit value inside an area surrounding the dumbbell and that sloped gradually to zero outside this region.  This absorbing boundary condition enabled the determination the atom loss as a function of time.

Figure~\ref{gpe_exp_compare_fig} compares the GP results with the data of Ref.\,\cite{nist_paper}.  The number imbalance, $\Delta N(t)$, is plotted versus time for three progressively wider (top to bottom, curves vertically offset for clarity) channels. As an aid to intuition, we introduce a Thomas--Fermi (TF) approximate channel width $w_{\mathrm{TF}}=((64/\pi)gn_{\mathrm{ch}}\omega_{z}/\omega_{y}^{3})^{1/4}$, (where $n_{\mathrm{ch}}$ is the number of atoms per unit length in the channel, see Supplementary Materials for details).  The channel widths range between 12 and 14 $\mu$m. The agreement between theory (shown in red), with no adjustable parameters, and experiment is good suggesting that the GP equation can accurately portray the behavior of average quantities such as $\Delta N(t)$.  It is therefore interesting to understand the GP--predicted transport dynamics and how it depends on channel shape.

The behavior of the number imbalance can be conveniently described in terms of a model RLC circuit with a time--dependent resistance, an initially charged capacitor and a switch that is closed at $t=0$. The capacitor charge ratio, $q(t)/q(0)$, in this circuit is equivalent to the number imbalance, $\Delta N(t)$,\,\cite{SciRep.3.1034} and the chemical capacitance, in analogy with the electronic capacitance, can be determined from the potential shape and the number of condensate atoms~\cite{SciRep.3.1034} (See Supplemental Material for a fuller description). 

Figure~\ref{gpe_deltaN}(a) displays typical behavior of $\Delta N(t)$ in which three distinct regimes of behavior are readily apparent. For $0 \le t \le t_{0}$ it drops rapidly from unity as the condensate flows down the channel and begins to fill the drain. We find that this behavior accurately matches a critically damped RLC circuit characterized by a capacitive discharge time, $\tau$. The optical density at several times during this period is shown in Fig.\ \ref{gpe_deltaN}(b).  At $t\approx t_{0}$, $\Delta N$ exhibits a ``kink'', that is an abrupt increase in the (negative) slope, after which the imbalance shows an approximately linear decrease between $t_{0} < t < t_{1}$.  At $t=t_{0}$, the resistance in the analog circuit abruptly increases and then begins a linear decrease to zero at $t=t_{1}$. After this, $\Delta N$ exhibits oscillations at frequency $\omega$ about a positive, but small, average value equivalent to resistanceless LC oscillations in the analog circuit.  The Supplemental Material contains details on this model circuit and how the capacitance and time--dependent resistance were calculated.

The presence of the kink can be understood from the top (blue) curve appearing in Fig.\ \ref{gpe_deltaN}(a).  This curve shows the ratio of the number of atoms located in the dumbbell area at time $t$ relative to the initial number.  It is clear from this curve that this ratio drops sharply at time $t \approx t_{0}$ indicating atom loss from the dumbbell area. This coincides with the appearance of the kink.  From the simulation images it can be seen that atoms leaving the dumbbell region area do so chiefly in the drain well. A sudden reduction of $N_{D}$, while keeping $N_{S}$ fixed, causes an increase in $\Delta N$ or, in this case, a slowing of its rate of decrease.

The question arises as to why the atoms are leaving the drain.  Atoms in the initial condensate (shown in the top picture in Fig.\ \ref{gpe_deltaN}(b)) obviously do not have enough energy (which is chiefly due to interactions) to jump out of the source well. When the early atoms arrive at the drain their energy is almost entirely kinetic but their total energy hasn't changed and they will still be unable to leave the drain.

One possible mechanism for atoms to gain the needed kinetic energy to escape the drain is through energy--redistributing collisions.  This is illustrated in Fig.\ \ref{gpe_deltaN}(c).  Atoms fan out as they exit the channel.  Some atoms flow straight along the dumbbell axis (straight--through atoms) while others diffract, bounce off the drain wall, and come back to the drain center (wall--bounce atoms).  Straight--through atoms colliding elastically with wall--bounce atoms can redistribute kinetic energy.  

In a classical collision of two identical particles of mass $m$, speed $v$, and colliding at an angle $\theta$, kinetic energy is transferred from one particle to the other after the collision.  The maximum possible kinetic energy of the more energetic particle is given by (see Supplemental Material for more details):
\begin{equation}
K_{\mathrm{f,max}} = K_{\mathrm{i}}\left(1 + \sin\theta\right).
\end{equation}
For the case illustrated in Fig.\ \ref{gpe_deltaN}, the angle measured from the simulation images is $\theta\approx 35^{\circ}$.  Assuming that the initial energy of colliding atoms was equal to the chemical potential of the initial condensate, for this case we have $K_{\mathrm{i}}/k\approx 65$\,nK and so $K_{\mathrm{f}}/k\approx 102$\,nK which is greater than the 97 nK well depth for this case.  

Thus even one collision can transfer enough kinetic energy to an atom to enable it to escape the drain.  Atom loss from the condensate is an additional dissipation mechanism, increasing the resistance already developed by the creation of vortices and solitons\,\cite{nist_paper}.  Atoms leaving the condensate presumably would enter the thermal cloud present in the experiment but not accounted for in the GP model.  In the experiment, condensate and thermal atoms cannot be distinguished making this loss process hard to detect.  

Finally we studied the dependence of the transport dynamics on the channel length and width by performing 252 simulations. Each simulation was characterized by a value of $\omega_{y}$, chosen from 21 values in the range $60\,\mathrm{Hz}\lesssim\omega_{y}/(2\pi)\lesssim 130\,\mathrm{Hz}$, and a value of $L_{c}$, chosen from 12 values in the range $2\,\mu\mathrm{m} \le L_{c} \le 24\,\mu\mathrm{m}$.  We calculated the number of atoms in the source, channel, and drain regions of the dumbbell potential as a function of time for a total simulation time of 0.5 seconds.  For each case we fit the number imbalance to find the values of the capacitive discharge time, $\tau$, and the LC oscillation frequency, $\omega$. 

\begin{figure}[thb]
\begin{center}
\includegraphics{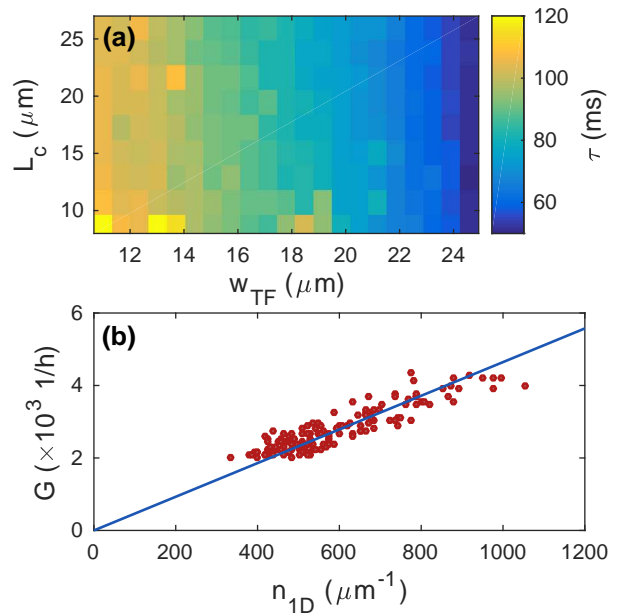}
\end{center}
\caption{(Color online) (a) Exponential decay time, $\tau(L_{c},w_{\mathrm{TF}})$, vs. the length of the channel $L_c$ and the Thomas--Fermi width of the channel, $w_{TF}$. (b) Conductance $G=1/R$ vs. one--dimensional channel density, $n_{1D}$, with a linear fit.}
\label{shape_study}
\end{figure}

The dependence of the exponential decay time, $\tau(L_{c},w_{\mathrm{TF}})$, on the channel length and width is shown in the density plot in Fig.\,\ref{shape_study}(a).  It is clear from this plot that $\tau = RC$ is independent of $L_{c}$.  Since we have fixed the number of condensate atoms and the shape of the wells, the capacitance, $C$, is also fixed.  The resistance is therefore independent of channel length suggesting a ``contact'' resistance in which dissapative processes giving rise to a resistance occur in the wells rather than in the channel.  Examples of such processes include atom loss and the formation of vortices and solitons.

The plot also shows that the decay time does depend on the channel width.  We averaged $\tau(L_{c},w_{\mathrm{TF}})$ over $L_{c}$ keeping $w_{\mathrm{TF}}$ fixed and found that the resulting average decay time, $\tau_{\mathrm{avg}}(w_{\mathrm{TF}})$, decayed according to a power law $w_{\mathrm{TF}}^{-1}$.  This differs from the Feynman model~\cite{feynman} where superfluid flow above a critical velocity from a channel into an infinite reservoir dissipates energy via the formation of a line of vortices.  The Feynman resistance varies inversely with the square of the channel width.  This difference may be due to the presence of other dissipation sources such as atom loss and the geometry of the well.  

Finally we analyzed the results of the study to determine how the channel resistance depends on the late--time 1D density of atoms in the channel, $n_{1D}$.  We used the fitted decay times averaged over all channel lengths for fixed channel width, $\tau_{\mathrm{avg}}(w_{\mathrm{TF}})$, to calculate the conductance, $G = 1/R = C/\tau_{\mathrm{avg}}(w_{\mathrm{TF}})$.  Figure\,\ref{shape_study}(b) shows a plot of this quantity versus $n_{1D}$.  The relationship between the conductance and $n_{1D}$ is linear as was seen in the experiment~\cite{nist_paper}.  The dependences of the oscillation frequency, $\omega$, on channel length, width, and channel 1D density were determined by fitting to power laws $L_{c}^{b_{l}}$,  $w_{\mathrm{TF}}^{b_{w}}$, and $n_{1D}^{b_{n}}$, respectively.  The results were $b_{l} = -0.18(2)$, $b_{w}=0.35(8)$, and $b_{n}=0.29(6)$. 

In conclusion, we studied the transport dynamics of a BEC confined in a quasi--planar, dumbbell atomtronic circuit.  We found that the results of GP--equation simulations matched the data of a recent experiment~\cite{nist_paper} given an accurate confining potential and absorbing boundary conditions.  The GP solution exhibited atom loss from the drain well, a phenomenon not easily detectable in the experiment.  We further proposed that this atom loss could be due to collisions that redistribute the atoms' kinetic energy.  Finally we presented a systematic study of the characteristics of the transport for a range of different channel lengths and widths.  

These results suggest that the operation of the class of atomtronic systems where a BEC is confined to a planar region with arbitrary in--plane potential can be simulated by using the GP equation.  We believe that the GP equation can be a useful tool in addressing the challenge of developing a theoretical framework for design of atomtronic circuits similar to the theoretical framework that already exists for designing electronic circuits.  If such a framework could be developed, it would enable atom circuits to be designed for applications not yet envisaged. 

\begin{acknowledgments}
This material is based upon work supported by the U.S.\ National Science Foundation under grant numbers PHY--1068761 and PHY--1413768, and the Physics Frontier Center @ JQI.  This work was also supported by a grant from the China Scholarship Council grant number [2012]3011.
\end{acknowledgments}

\bibliography{dumbell_paper}{}

\widetext
\clearpage
\begin{center}
\textbf{\large Supplemental Material: Superfluid transport dynamics in a capacitive atomtronic circuit}
\end{center}
\setcounter{equation}{0}
\setcounter{figure}{0}
\setcounter{table}{0}
\setcounter{page}{1}
\makeatletter

In this Supplemental Material section we provide extra material on the following topics: (1) a description of how the potential parameters used to model the NIST experiment were determined, (2) an explanation of the fitting procedure used in the systematic study of transport behavior for different channel shapes, (3) the details of the variable--resistance RLC circuit model, (4) a derivation of the model capacitance of the dumbbell circuit, (5) a derivation of the Thomas--Fermi channel width, $w_{\mathrm{TF}}$, and (6) a derivation of the formula for maximum kinetic energy transfer in a classical, elastic collision.

\section{Determination of dumbbell potential parameters}
\label{dumbbell_parameters}

We reiterate the potential here for convenience.  The full model dumbbell potential consists of sheet, well, and channel potentials:
\begin{eqnarray}
V_{\rm trap}({\bf r},t) 
&=& 
\frac{1}{2}M
\left(
\omega_{\rm sh,x}^{2}x^{2} + 
\omega_{\rm sh,y}^{2}y^{2} +
\omega_{\rm sh,z}^{2}z^{2}
\right)\nonumber\\
&+&
V_{\rm well}(x,y) + 
V_{\rm channel}(x,y)
\end{eqnarray}
The $z$ axis is vertical, the $x$ axis lies along the line joining the two well centers, and the $y$ axis is perpendicular to the channel. There is also a gate potential but it was modeled as a high step located along the $y$ axis and was present only during formation of the initial condensate we neglect it here. 

The model well and channel potentials have the form
\begin{eqnarray}
V_{\rm well}(x,y)
&=& 
V_{d}\sum_{k=1,2}\frac{1}{2}
\left[1+\tanh\left(\frac{\rho_{k}(x,y)-r_{k}}{b}\right)\right],\nonumber\\
\rho_{k}(x,y)
&\equiv&
\sqrt{(x-x_{k})^{2}+(y-y_{k})^{2}},
\quad
k = 1,2,\nonumber\\
V_{\rm channel}(x,y) 
&=& 
V_{\rm step}+
\frac{1}{2}M\omega_{y}^{2}y^{2}.
\end{eqnarray}
The channel potential has two circular wells having centers $(x_{k},y_{k})$ and radii $r_{k}$ where $k=1,2$ and depth $V_{d}$.  The hardness of the well edges was varied by adjusting the value of $b$ which is the range over which the step rises from zero to one.  The channel potential is harmonic along the $y$ direction (due to instrument resolution~\cite{nist_paper}) plus a step, $V_{\rm step}$.

The full mask potential is equal to $\min(V_{\rm well},V_{\rm channel})$ between the wells ($-r_{1}\le x \le r_{2}$) and equal to $V_{\rm well}$ outside this region. The gate potential is a high step function parallel to the $y$ axis and located in the center of the channel.  The sheet--potential frequencies were determined to be $\omega_{\rm sh,x}/2\pi=10$ Hz, $\omega_{\rm sh,y}/2\pi\approx 0$ Hz, and $\omega_{\rm sh,z}/2\pi=529$ Hz. The well radii were $r_{1}=r_{2}= 24\ \mu$m, the well depth was $V_{d}=83$ nK, and the hardness parameter was $b=0.2\ \mu$m.  The well centers were separated by $74\ \mu$m. In the channel $V_{\rm step}\approx 20$ nK and $\omega_{y}/2\pi$ varied between 110 and 130 Hz. 

Because of imaging aberrations, the exact channel potential is unknown and cannot be determined {\it a priori}.  We therefore chose to model the channel potential in the simple way described above, using the combination of a harmonic trapping potential $\omega_y\propto \sqrt{V_d}$ and a $V_{\rm step}$ that is independent of $V_d$.  With two free parameters -- the proportionality constant between $\omega_y$ and $\sqrt{V_d}$ and $V_{\rm step}$ -- we found we could accurately predict the measured 1D equilibrium densities in the channel.  Other observables like the Thomas-Fermi width or the 2-D density of atoms are compromised by the aberrations; as such, the 1-D density is the only reliable measure by which we can model the channel potential.  The reservoir potentials, which are much larger than the channel, are less affected by the imaging aberrations and thus we found parameters ($\omega_{{\rm sh},x}$ and $\omega_{{\rm sh},y}$, $b$, and $V_d$) that best reproduced the measured 2-D density.  

\section{Shape--study fitting procedure}
\label{shape_study_fit}

The study of the transport dynamics of condensate released into the dumbbell potential across a range of different channel lengths and widths described in the main text consisted of two phases.  The first phase was the simulation of the condensate behavior using the Gross--Pitaevskii (GP) equation for each of the 252 cases of fixed channel length and width. For each case, the number of atoms in the source well, channel, and drain well as a function of elapsed time following condensate was calculated and saved. 
In the second phase, for each channel shape, the number imbalance, $\Delta N(t)$, was calculated and a fit was performed to extract the capacitive decay constant, $\tau$, and the LC oscillation frequency, $\omega$.

The simulations performed in the first phase were done by numerical solution of the GP equation.  The procedure for this was the same as described in the main text for the experimental simulations.  The hybrid Lagrangian variational equations (HLVM) of motion for the 3D GP were solved using the split--step, Crank--Nicolson algorithm under conditions of space and time step size the same as for the experiments.  Each simulation produced a time--tagged wave function, stored on a 2D space grid, at 400 equally spaced times during the interval $0 \le t \le 0.5$ seconds and these wave functions were immediately used to compute the populations (numbers of atoms) in the source, channel, and drain regions.  These numbers were stored and the wave functions were not retained.  

In order to calculate the population in a particular region, we integrated squared modulus of the HLVM wave function over the given region.  The HLVM wave function (up to an irrelevant phase factor) has the form~\cite{PhysRevE.86.056710}:
\begin{equation}
\Psi(x,y,z,t) = 
\left(
\frac{1}{\sqrt{\pi}w(t)}
\right)^{1/2}
\exp\left\{-\frac{z^{2}}{2w^{2}(t)}\right\}
\psi(x,y,t).
\end{equation}
The number of atoms in the source region at time $t$ is given by
\begin{eqnarray}
N_{S}(t) 
&=& 
\int\limits_{\substack{3D\ source\\region}}\,d^{3}r
\left|\Psi(x,y,z,t)\right|^{2} =
\left(
\frac{1}{\sqrt{\pi}w(t)}
\right)
\int_{-\infty}^{+\infty}dz\,
\exp\left\{-\frac{z^{2}}{w^{2}(t)}\right\}
\iint\limits_{\substack{2D\ source\\region}}dx\,dy
\left|\psi(x,y,t)\right|^{2}\nonumber\\
N_{S}(t) 
&=& 
\iint\limits_{\substack{2D\ source\\region}}dx\,dy
\left|\psi(x,y,t)\right|^{2},
\end{eqnarray}
and similarly for the channel and drain regions.  The source, channel, and drain populations were computed numerically at each time during a given simulation and the result was stored.  This was done for all 252 channel--shape cases.

In phase two of the shape study, the time--dependent number imbalance, $\Delta N(t) = (N_{S}(t)-N_{D}(t))/(N_{S}(t)+N_{D})$, was fitted with a function that enabled the estimation of the decay constant, $\tau$, and the LC oscillation frequency, $\omega$.
The fitting function had the following form:
\begin{equation}
\Delta N(t) = 
\left[
\frac
{1}
{1+
\exp
\left\{
\frac
{t-t_{c}}
{t_{w}}
\right\}
}
\right]
e^{-t/\tau} + 
\left[
\frac
{1}
{1+
\exp
\left\{
-\frac
{t-t_{c}}
{t_{w}}
\right\}
}
\right]
\left(
b\cos\left(\frac{2\pi t}{T}+\phi\right) + c
\right).
\end{equation}
This function only assumes capacitive discharge, $e^{-t/t_{F}}$, and LC--oscillation behavior, $b\cos\left(\frac{2\pi t}{T}+\phi\right) + c$, and consists of these functions multiplied by turn--off and a turn--on functions, respectively. These turn--off/turn--on functions are set off in square brackets in the above equation.  The fit parameters are $t_{c}$, $t_{w}$, $\tau$, $b$, $T$, $\phi$, and $c$.  The decay constant, $\tau$, is already one of the fit parameters and the LC oscillation frequency can be calculated by $\omega = 2\pi/T$.  

The number imbalance associated with each channel--shape simulation was fit with this function to get $\tau$ and $\omega$.  These results were then, themselves, analyzed for dependence on channel length and width as described in the main text.

\section{Variable resistance RLC circuit model}
\label{RLC_model}

\begin{figure*}[htb]
\begin{center}
\includegraphics[width=6.75in]{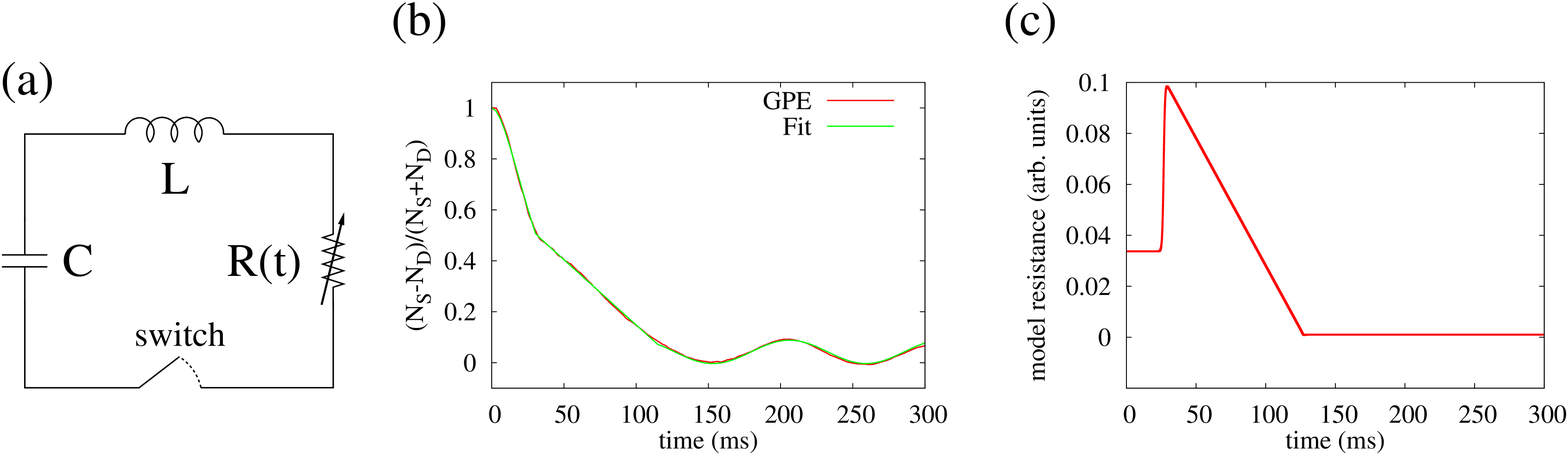}
\end{center}
\caption{(color online) (a) Model RLC circuit with variable resistance. (b) Number imbalance, $\Delta N(t)$, from the Gross--Pitaevskii (GP) equation and fitted result (see text) for a dumbbell potential with channel length $L_{c}=20\,\mu\mathrm{m}$ and and Thomas--Fermi width of $w_{\mathrm{TF}}\approx 22\,\mu\mathrm{m}$. (c) The time--dependent resistance, $R(t)$, derived from the fitted number imbalance.}
\label{variable_rcl}
\end{figure*}

In this section we give more details about the variable--resistance RLC circuit model described in the main paper.  This circuit model consists of an initially charged capacitor of capacitance $C$, an inductor with inductance $L$, an initially open switch, and a resistor with time--dependent resistance, $R(t)$.  This circuit is shown in Fig.\ \ref{variable_rcl}(a).

The goal of this modeling exercise is to find out if such a model circuit with some variable resistance, $R(t)$, is capable of reproducing the behavior of the number imbalance, $\Delta N(t)$, of the dumbbell circuit as produced by the GP equation.  Thus we need to find the functional form of $R(t)$.  The method for determining it will be to equate the normalized charge on the capacitor, 
\begin{equation}
\bar{q}(t)\equiv q(t)/q(0), 
\end{equation}
where $q(t)$ is the capacitor charge in the model circuit, to the number imbalance.  Next we find a formula for $R(t)$ in terms of $\bar{q}(t)$ using Kirchhoff's circuit rules.  Finally we find a fit to $\bar{q}(t)$ and use this fitting function to get $R(t)$. 

We can apply Kirchhoff's rule to the circuit in Fig.\ \ref{variable_rcl}(a) where we assume an instantaneous current $i(t)$ flowing in the clockwise direction and an instantaneous charge $q(t)$ on the capacitor.  Then we have the following two equations:
\begin{eqnarray}
L\frac{di}{dt} + R(t)i(t) + \frac{1}{C}q(t) &=& 0\nonumber\\
\frac{dq}{dt} &=& i(t)
\end{eqnarray}
and we can combine these yielding one equation for $q(t)$:
\begin{eqnarray}
L\frac{d^{2}q}{dt^{2}} + R(t)\frac{dq}{dt} + \frac{1}{C}q(t) &=& 0\nonumber\\
R(t) = -
\left(
\frac{\ddot{\bar{q}}(t)+\omega^{2}\bar{q}(t)}{C\omega^{2}\dot{\bar{q}}}
\right)
\end{eqnarray}
where $\omega\equiv 1/\sqrt{LC}$ is the frequency of LC oscillations that the circuit would have if $R=0$ and we have written the result in terms of $\bar{q}$.  The values of $q(t)$ and $\omega$ will be found by fitting. The value of the capacitance can be calculated given the dumbbell potential and the number of condensate atoms and so is assumed to be known here.

The number imbalance has distinct behavior during Interval (a): $0 \le t \le t_{0}$, Interval (b): $t_{0} \le t \le t_{1}$ and Interval (c): $t > t_{1}$ as explained in the main text.  Thus we chose a different function for $\Delta N(t)$ on each interval.  Our fitting function was the following:
\begin{equation}
\Delta N(t) = \frac{q(t)}{q(0)} \equiv \bar{q}(t) = 
\left\{
\begin{array}{lc}
(1+\omega_{1}t)e^{-\omega_{1}t} & 0 \le t \le t_{0}\\
q_{c} - I_{c}(t - t_{0}) & t_{0} \le t \le t_{1}\\
b\cos(\omega_{2}t + \phi) + c & t < t_{1}
\end{array}
\right\}.
\end{equation}
We fit the frequencies appearing in Intervals (a) and (c) separately.  However, we found that these frequencies were nearly the same in most cases. The fitting parameters were $\omega_{1}$, $q_{c}$, $I_{c}$, $b$, $\omega_{2}$, $\phi$, $c$, $t_{0}$, and $t_{1}$. Thus, for the final value of the fitting function, we set $\omega_{1} = \omega_{2} = (\omega_{1}+\omega_{2})/2$.  

An example of the result of this fitting procedure is shown in Fig.\ \ref{variable_rcl}(b). The graph shows a plot of number imbalance for a dumbbell potential with channel length $L_{c} = 20\,\mu\mathrm{m}$ and TF width $w_{\mathrm{TF}}=22\,\mu\mathrm{m}$ (red curve) along with the result of the fit (green curve).  The fit is very good as would be expected given the number of fitting parameters.

Finally, this fitting function can be used to find the time--dependent resistance, $R(t)$, as already described.  This quantity is plotted in Fig.\ \ref{variable_rcl}(c).  The plot shows that $R$ is essentially constant during $0 \le t \le t_{0}$.  At $t=t_{0}$, where $\Delta N$ displays a kink, the resistance abruptly increases after which is decreases linearly to zero at $t=t_{1}$ and remains zero thereafter.  A summary of this behavior appears in the main text.

\section{Calculation of the capacitance of the dumbbell--potential BEC}
\label{TF_cyl_mu}

\subsection{Definition of chemical capacitance}
Here we derive the capacitance of the BEC in the dumbbell potential.  The chemical capacitance is defined in analogy with the definition for an electronic capacitance.  For the electronic capacitor, if an external agent moves a (positive) charge $\delta q$ from one initially neutral plate (drain plate) to the other plate (source plate), then the source plate has a net charge of $+\delta q$ while the drain plate has a net charge $-\delta q$.  The source reservoir gets its name from the fact that moving ``charges'' (either real positive charges or atoms) will initially flow from source to drain to regain equilibrium.  The charge difference (source charge - drain charge) is $+2\delta q$ however the {\em charge on the electronic capacitor} is regarded as being $+\delta q$.  Once there is a charge on the capacitor, it causes a voltage difference, $\delta V$, to develop and the electronic capacitance is defined as
\begin{equation}
C_{\mathrm{elec}} \equiv \frac{\delta q}{\delta V}.
\end{equation}
This motivates the definition of the chemical capacitance.

The chemical capacitance is defined analogously. If the total number of atoms is $N$ and there are equal numbers of atoms ($N/2$) in each well (this is equilibrium) and the external agent moves $\delta N$ atoms from the drain well to the source well, then the number of atoms in each well is
\begin{equation}
N_{S} = \frac{1}{2}N + \delta N
\quad\mathrm{and}\quad
N_{D} = \frac{1}{2}N - \delta N.
\end{equation}
Analogous to the electronic capacitor the difference in the number of atoms between the wells is $+2\delta N$. However, we shall regard the {\em charge on the chemical capacitor} as being $\delta N$ atoms.  Once there is a charge on the chemical capacitor, it causes a chemical potential difference, $\delta\mu$, to develop and the chemical capacitance is defined as
\begin{equation}
C_{\mathrm{chem}} \equiv \frac{\delta N}{\delta\mu}.
\end{equation}
We can find an approximate expression for this based on the TF approximation for the chemical potential of a BEC in a cylindrical well potential.

\subsection{The chemical potential for a BEC in a cylindrical well}
Consider a BEC having $N$ atoms confined in a cylindrical well potential. To get an expression for the chemical potential we will assume that the Thomas--Fermi approximation applies and that the BEC is confined radially in a hard--walled cylindrical well of radius $R$ and axially in a harmonic potential.  We assume that the confining potential is therefore (using cylindrical coordinates $(r,\theta,z)$
\begin{equation}
V_{\mathrm{trap}}(r,\theta,z) = \frac{1}{2}M\omega_{z}^{2}z^{2} + V_{\mathrm{well}}(r)
\end{equation}
where
\begin{equation}
V_{\mathrm{well}}(r) = 
\left\{
\begin{array}{lc}
0 & 0 \le r \le R\\
\infty & R < r < \infty
\end{array}
\right\}.
\end{equation}
In the potential actually present in the experiment, there is also a weak harmonic confinement along the $x$ direction whose zero is offset from the origin (assumed here to be at the center of the source well).  The effect of this extra harmonic potential is to produce a small but noticeable gradient along the $x$ direction in the initial condensate density profile.  Since this extra potential is so weak ($\omega_{\mathrm{sh,x}}/2\pi\approx 10\mathrm{ Hz}$), it does not appreciably change the value of the chemical potential or its dependence on $N$.

The condensate wave function, $\psi_{0}(\mathbf{r})$, satisfies the time--independent Gross--Pitaevskii (GP) equation:
\begin{equation}
-\frac{\hbar^{2}}{2M}\nabla^{2}\psi_{0}(\mathbf{r}) +
V_{\mathrm{trap}}(r,\theta,z)\psi_{0}(\mathbf{r}) +
gN|\psi_{0}(\mathbf{r})|^{2}\psi_{0}(\mathbf{r}) = 
\mu_{0}\psi_{0}(\mathbf{r}).
\end{equation}
where $M$ is the mass of a condensate atom, $N$ is the number of condensate atoms, $g = 4\pi\hbar^{2}a_{s}/M$ is the strength of the binary scattering of condensate atoms, $a_{s}$ is the $s$--wave scattering length, and $\mu_{0}$ is the chemical potential of the condensate.  The chemical potential is the energy required to add another atom to the condensate.

We will derive an approximate expression for the chemical potential by using the Thomas--Fermi (TF) approximation.  The TF approximation is valid whenever the interaction energy is much larger than the kinetic energy.  When this is the case we have $\psi_{0}(\mathbf{r})\approx\psi_{0}^{(\mathrm{TF})}(\mathbf{r})$, $\mu_{0}\approx\mu_{0}^{(\mathrm{TF)}}$, and these TF--approximate quantities satisfy the GP equation where the kinetic--energy term is neglected:
\begin{equation}
V_{\mathrm{trap}}(r,\theta,z)\psi_{0}^{(\mathrm{TF)}}(\mathbf{r}) +
gN|\psi_{0}^{(\mathrm{TF)}}(\mathbf{r})|^{2}\psi_{0}^{(\mathrm{TF)}}(\mathbf{r}) = 
\mu_{0}^{(\mathrm{TF)}}\psi_{0}^{(\mathrm{TF)}}(\mathbf{r}).
\end{equation}
The formal solution of this equation is
\begin{equation}
gN|\psi_{0}^{(\mathrm{TF)}}(\mathbf{r})|^{2} = 
\left\{
\begin{array}{lc}
\mu_{0}^{(\mathrm{TF)}}-V_{\mathrm{trap}}(r,\theta,z) &
\mu_{0}^{(\mathrm{TF)}}-V_{\mathrm{trap}}(r,\theta,z) \ge 0\\
0 & \quad\mathrm{otherwise}\quad
\end{array}
\right\}
\end{equation}
We can now insert the particular form of the potential into the above to get 
\begin{equation}
gN|\psi_{0}^{(\mathrm{TF)}}(\mathbf{r})|^{2} = 
\left\{
\begin{array}{lc}
\frac{1}{2}M\omega_{z}^{2}z_{\mathrm{TF}}^{2} -
\frac{1}{2}M\omega_{z}^{2}z^{2} &
0 \le r \le R,\ \mathrm{and}\ |z| \le z_{\mathrm{TF}}\\
0 & \quad\mathrm{otherwise}\quad
\end{array}
\right\}
\end{equation}
Where we have defined the TF $z$ radius as
\begin{equation}
\mu_{0}^{(\mathrm{TF)}} \equiv 
\frac{1}{2}M\omega_{z}^{2}z_{\mathrm{TF}}^{2}.
\end{equation}
The values of $z_{\mathrm{TF}}$ and thus $\mu_{0}^{(\mathrm{TF)}}$ are found by requiring that the TF wave function be normalized to unity or, more conveniently
\begin{equation}
gN = gN\int\,d^{3}r|\psi_{0}^{(\mathrm{TF)}}(\mathbf{r})|^{2}.
\end{equation}
Thus we have
\begin{eqnarray}
gN 
&=&
\int_{0}^{2\pi}\,d\theta\,
\int_{-z_{\mathrm{TF}}}^{+z_{\mathrm{TF}}}\,dz\,
\int_{0}^{R}\,rdr\,
\frac{1}{2}M\omega_{z}^{2}
\left(z_{\mathrm{TF}}^{2} - z^{2}\right)\nonumber\\
&=&
\left(\frac{1}{2}M\omega_{z}^{2}\right)
(2\pi)\left(\frac{1}{2}R^{2}\right)
\left(\frac{4}{3}z_{\mathrm{TF}}^{3}\right)\nonumber\\
&=&
\left(\frac{4}{3}\pi R^{2}\right)
\frac
{
\left(\mu_{0}^{(\mathrm{TF)}}\right)^{3/2}
}
{
\left(\frac{1}{2}M\omega_{z}^{2}\right)^{1/2}
}.
\end{eqnarray}
So the expression for the TF chemical potential is
\begin{equation}
\mu_{0}^{(\mathrm{TF)}} = 
\left(
\frac
{
(gN)\left(\frac{1}{2}M\omega_{z}^{2}\right)^{1/2}
}
{
\left(\frac{4}{3}\pi R^{2}\right)
}
\right)^{2/3} \equiv \alpha N^{2/3}
\label{mu_cyl_tf}
\end{equation}

\subsection{The chemical capacitance in the Thomas--Fermi approximation}
Now consider the chemical potential difference between the source well having $N_{S}$ atoms and the drain well with $N_{D}$ atoms.  Using Eq.\ (\ref{mu_cyl_tf}) we have 
\begin{eqnarray}
\delta\mu &=& 
\mu(N_{S}) - \mu(N_{D}) = \alpha N_{S}^{2/3} - \alpha N_{D}^{2/3}\nonumber\\
&=& 
\alpha\left(\frac{1}{2}N + \delta N\right)^{2/3} - 
\alpha\left(\frac{1}{2}N - \delta N\right)^{2/3}\nonumber\\
&=&
\alpha\left(\frac{1}{2}N\right)^{2/3}
\left[
\left(1 + \frac{\delta N}{\tfrac{1}{2}N}\right)^{2/3} -
\left(1 - \frac{\delta N}{\tfrac{1}{2}N}\right)^{2/3}
\right]\nonumber\\
&\approx&
\alpha\left(\frac{1}{2}N\right)^{2/3}
\left[
\left(1 + \frac{2}{3}\frac{\delta N}{\tfrac{1}{2}N}\right) -
\left(1 - \frac{2}{3}\frac{\delta N}{\tfrac{1}{2}N}\right)
\right]\nonumber\\
\delta\mu 
&=& 
\alpha\left(\frac{1}{2}N\right)^{2/3}
\frac{4}{3}\frac{\delta N}{\tfrac{1}{2}N} = 
\frac{4\alpha\delta N}{3\left(\tfrac{1}{2}N\right)^{1/3}}
\end{eqnarray}
where we assumed that $\delta N \ll \tfrac{1}{2}N$ in the above derivation.  This linear approximation is good to 10\% over a number imbalance up to $\delta N/(\tfrac{1}{2}N) = \pm 0.93$~\cite{nist_paper}.

Finally we can get an approximate expression for the chemical capacitance:
\begin{equation}
C_{\mathrm{chem}} = 
\frac{\delta N}{\delta\mu} = 
\frac{3\left(\tfrac{1}{2}N\right)^{1/3}}{4\alpha}.
\end{equation}
This expression for $C_{\mathrm{chem}}$ only depends on the shape of the potential and the total number of condensate atoms. 

\section{The Thomas--Fermi channel width}
\label{TF_width}

In this section we provide a derivation of the Thomas--Fermi condensate width in the dumbbell potential.  This width enables us to define the channel shape in a more intuitive way.  The channel is assumed to have a length $L_{c}$ along the line joining the two wells of the dumbbell ($x$ axis) with hard walls located at $x = \pm L_{c}/2$.  In between these walls, the potential is assumed to be harmonic in $y$ and $z$ plus a constant step:
\begin{equation}
V_{\mathrm{ch}}(x,y,z) =
\left\{
\begin{array}{lc}
V_{\mathrm{step}} + 
\frac{1}{2}M\omega_{y}^{2}y^{2} +
\frac{1}{2}M\omega_{z}^{2}z^{2} & 
|x| \le \tfrac{1}{2}L_{c}\\
\infty & |x| > \tfrac{1}{2}L_{c}
\end{array}
\right\}
\end{equation}

The solution of the time--independent Gross--Pitaevskii (GP) equation, $\psi$, can be approximated by the Thomas--Fermi solution $\psi_{\mathrm{TF}}$ where the kinetic--energy term in the GP is neglected:
\begin{eqnarray}
-\frac{\hbar^{2}}{2M}\nabla^{2}\psi(\mathbf{r}) + 
V_{\mathrm{ch}}(\mathbf{r})\psi(\mathbf{r}) +
gN_{\mathrm{ch}}|\psi|^{2}\psi(\mathbf{r})
&=& \mu\psi(\mathbf{r})\nonumber\\
V_{\mathrm{ch}}(\mathbf{r})\psi_{\mathrm{TF}}(\mathbf{r}) +
gN_{\mathrm{ch}}|\psi_{\mathrm{TF}}(\mathbf{r})|^{2}\psi_{\mathrm{TF}}(\mathbf{r})
&=&
\mu_{\mathrm{TF}}\psi_{\mathrm{TF}}(\mathbf{r})\nonumber\\
\end{eqnarray}
where we have assumed that there are $N_{\mathrm{ch}}$ atoms in the channel. The Thomas--Fermi--approximate solution can be written as
\begin{eqnarray}
gN_{\mathrm{ch}}|\psi_{\mathrm{TF}}(\mathbf{r})|^{2} 
&=&
\left\{
\begin{array}{cc}
\mu_{\mathrm{TF}} - V_{\mathrm{ch}}(\mathbf{r}) & 
\mu_{\mathrm{TF}} \ge V_{\mathrm{ch}}(\mathbf{r})\\
0 & \mathrm{otherwise}
\end{array}
\right\}\nonumber\\
&=&
\left\{
\begin{array}{cc}
\mu_{\mathrm{TF}} - 
\left(V_{\mathrm{step}} + 
\frac{1}{2}M\omega_{y}^{2}y^{2} +
\frac{1}{2}M\omega_{z}^{2}z^{2}\right) & 
\mu_{\mathrm{TF}} \ge V_{\mathrm{ch}}(\mathbf{r})\\
0 & \mathrm{otherwise}
\end{array}
\right\}\nonumber\\
\end{eqnarray}
The value of $\mu_{\mathrm{TF}}$ is determined by normalization.

Thus we require that
\begin{equation}
\int\,d^{3}r\,|\psi_{\mathrm{TF}}(\mathbf{r})|^{2} = 1
\quad\mathrm{or,\ equivalently,}\quad
\int\,d^{3}r\,gN_{\mathrm{ch}}|\psi_{\mathrm{TF}}(\mathbf{r})|^{2} = gN_{\mathrm{ch}}.
\end{equation}
Using the explicit TF solution we have
\begin{eqnarray}
gN_{\mathrm{ch}}
&=&
\int\,d^{3}r\,gN_{\mathrm{ch}}|\psi_{\mathrm{TF}}(\mathbf{r})|^{2}\nonumber\\
&=&
\int_{-L{c}/2}^{+L_{c}/2}dx\,
\int_{-y_{\mathrm{TF}}}^{+y_{\mathrm{TF}}}dy\,
\int_{-z_{\mathrm{TF}}(y)}^{+z_{\mathrm{TF}}(y)}dz\,
\left[
(\mu_{\mathrm{TF}} - 
V_{\mathrm{step}}) - 
\frac{1}{2}M\omega_{y}^{2}y^{2} -
\frac{1}{2}M\omega_{z}^{2}z^{2}
\right]\nonumber\\
\end{eqnarray}
where $z_{\mathrm{TF}}(y)$ is satisfies
\begin{equation}
\frac{1}{2}M\omega_{z}^{2}z_{\mathrm{TF}}^{2}(y) =
(\mu_{\mathrm{TF}} - 
V_{\mathrm{step}}) - 
\frac{1}{2}M\omega_{y}^{2}y^{2}.
\end{equation}
Performing the integral over $z$ gives us
\begin{eqnarray}
gN_{\mathrm{ch}}
&=&
\frac{4}{3}\left(\frac{1}{2}M\omega_{z}^{2}\right)
\int_{-L{c}/2}^{+L_{c}/2}dx\,
\int_{-y_{\mathrm{TF}}}^{+y_{\mathrm{TF}}}dy\,
\left(
\frac
{(\mu_{\mathrm{TF}} - 
V_{\mathrm{step}}) - 
\frac{1}{2}M\omega_{y}^{2}y^{2}}
{\tfrac{1}{2}M\omega_{z}^{2}}
\right)^{3/2}
\label{int_over_z}
\end{eqnarray}
Now $y_{\mathrm{TF}}$ is the edge where the $y$ integrand goes to zero and satisfies the following condition:
\begin{equation}
\frac{1}{2}M\omega_{y}^{2}y_{\mathrm{TF}}^{2} = 
\mu_{\mathrm{TF}} - 
V_{\mathrm{step}}
\end{equation}
This expression enables us to define the Thomas--Fermi width, $w_{\mathrm{TF}}$, as follows:
\begin{equation}
w_{\mathrm{TF}} = 
2y_{\mathrm{TF}} = 
2
\left(
\frac{
\mu_{\mathrm{TF}} - 
V_{\mathrm{step}}
}
{
\tfrac{1}{2}M\omega_{y}^{2}
}
\right)^{1/2}
\end{equation}
So now we need to find $\mu_{\mathrm{TF}} - V_{\mathrm{step}}$ by evaluating the rest of the integral in Eq.\ (\ref{int_over_z}).  The result is
\begin{eqnarray}
gN_{\mathrm{ch}}
&=&
\left(\frac{\pi}{2}\right)
y_{\mathrm{TF}}^{4}
\frac
{
\left(\frac{1}{2}M\omega_{y}^{2}\right)^{3/2}
}
{
\left(\frac{1}{2}M\omega_{z}^{2}\right)^{1/2}
}
L_{c},
\end{eqnarray}
where
\begin{equation}
y_{\mathrm{TF}}^{4} = 
\left(
\frac{\mu_{\mathrm{TF}}-V_{\mathrm{step}}}
{\tfrac{1}{2}M\omega_{y}^{2}}
\right)^{2}.
\end{equation}
Inserting this into the equation for $gN_{\mathrm{ch}}$ gives us an expression from which we can evaluate $\mu_{\mathrm{TF}}-V_{\mathrm{step}}$:
\begin{eqnarray}
gN_{\mathrm{ch}}
&=&
\left(\frac{\pi}{2}\right)
\left(
\frac{\mu_{\mathrm{TF}}-V_{\mathrm{step}}}
{\tfrac{1}{2}M\omega_{y}^{2}}
\right)^{2}
\frac
{
\left(\frac{1}{2}M\omega_{y}^{2}\right)^{3/2}
}
{
\left(\frac{1}{2}M\omega_{z}^{2}\right)^{1/2}
}
L_{c} =
\frac
{
\tfrac{1}{2}\pi L_{c}
}
{
\left(\tfrac{1}{2}M\omega_{y}^{2}\right)^{1/2}
\left(\tfrac{1}{2}M\omega_{z}^{2}\right)^{1/2}
}
\left(
\mu_{\mathrm{TF}}-V_{\mathrm{step}}
\right)^{2}
\end{eqnarray}
We can use this equation to solve for $\mu_{\mathrm{TF}}-V_{\mathrm{step}}$:
\begin{equation}
\mu_{\mathrm{TF}}-V_{\mathrm{step}} = 
\left(
\frac
{
\left(gN_{\mathrm{ch}}\right)
\left(\tfrac{1}{2}M\omega_{y}^{2}\right)^{1/2}
\left(\tfrac{1}{2}M\omega_{z}^{2}\right)^{1/2}
}
{
\tfrac{1}{2}\pi L_{c}
}
\right)^{1/2}.
\end{equation}
With this we can now get the final result for $w_{\mathrm{TF}}$:
\begin{equation}
w_{\mathrm{TF}} = 2y_{\mathrm{TF}} = 
2
\left[
\left(\frac{2}{\pi}\right)
\frac
{
\left(gN_{\mathrm{ch}}\right)
\left(\frac{1}{2}M\omega_{z}^{2}\right)^{1/2}
}
{
L_{c}\left(\frac{1}{2}M\omega_{y}^{2}\right)^{3/2}
}
\right]^{1/4}
\end{equation}
The channel width can be obtained from this expression for each of the simulations in the channel--shape study by using the simulation value of $N_{\mathrm{ch}}$.  We compared the predictions of this formula with the channel widths as determined by inspection of the images at the end of each simulation and found good agreement between them.  This formula enables us to present the channel--shape study results in terms of channel length and width.  This is more intuitive in terms of the channel length and transverse harmonic frequency, $\omega_{y}$. 

\section{Kinetic energy transfer in a classical collision}
\label{KE_transfer}

In this section we derive a formula for the final kinetic energies of two identical particles that collide at an angle $\theta$.  This result enables us to estimate the amount of kinetic energy that can be transferred in a collision of wall--bounce atoms in the drain well with atoms that flow directly into the well.  This estimate can be compared with the depth of the well to determine if this kinetic--energy transfer is a viable mechanism for atom loss from the dumbbell region.

\begin{figure}[H] 
\begin{center}
\raisebox{-0.5\height}
{\includegraphics[width=0.5\linewidth]{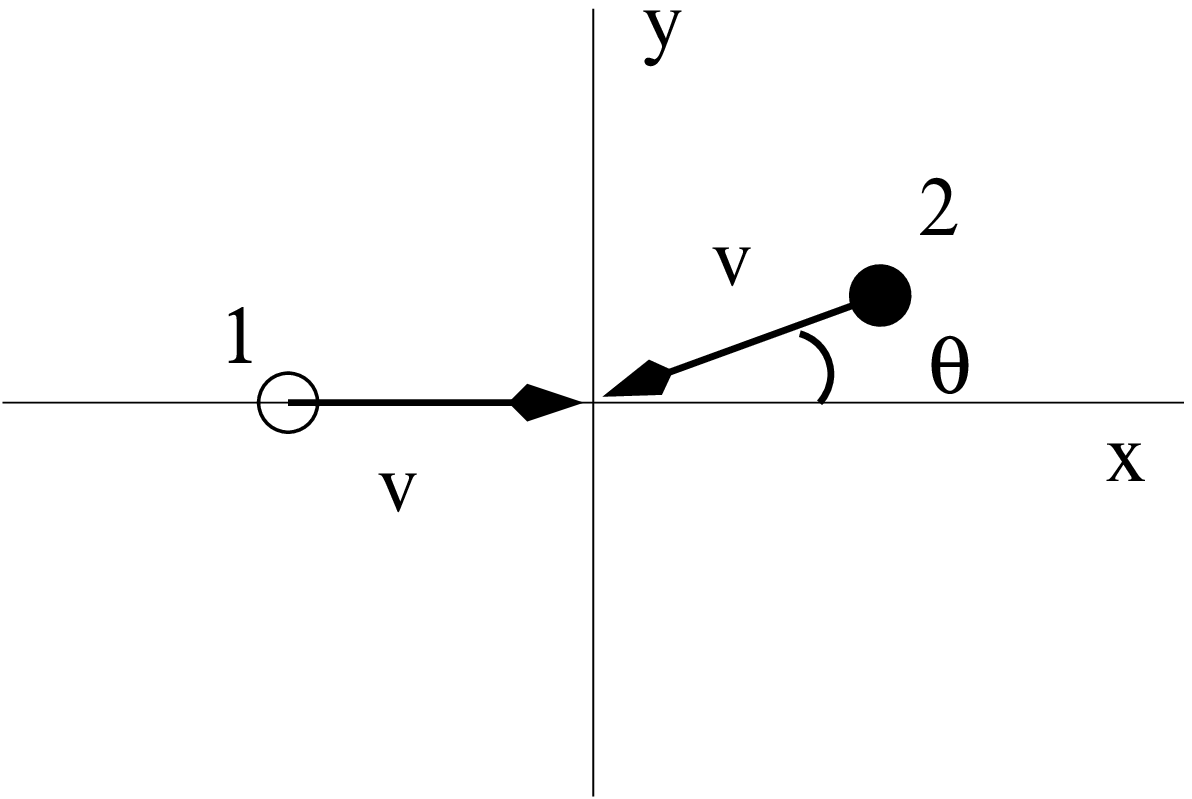}}
\end{center}
\caption{Classical model for a collision of atoms that bounce off the wall with those that come straight through.  Two particles each having mass $m$ and speed $v$ collide at an angle $\theta$.}
\label{colliding_atoms}
\end{figure}

We assume first that the colliding atoms have the same initial kinetic energy and that this kinetic energy comes entirely from the interaction energy of the initially confined condensate in the source well.  In the source well, almost all of the particle energy is interaction energy since the potential energy and kinetic energies are (nearly) zero.  After release, particles that flow into the channel have all of their initial interaction energy converted to kinetic energy.

To develop a classical model we imagine two identical particles of mass $M$ and initial speed $v$ colliding elastically at an angle $\theta$ as shown in Fig.\ \ref{colliding_atoms}. The central question is: how much kinetic energy can be transferred from one particle to the other in this perfectly elastic, momentum--conserving collision? We will impose these conservation laws on the collision in the center--of--mass (CM) frame.  These lab and CM frames are illustrated in the following figure:
\begin{figure}[H] 
\begin{center}
\includegraphics[width=0.75\linewidth]{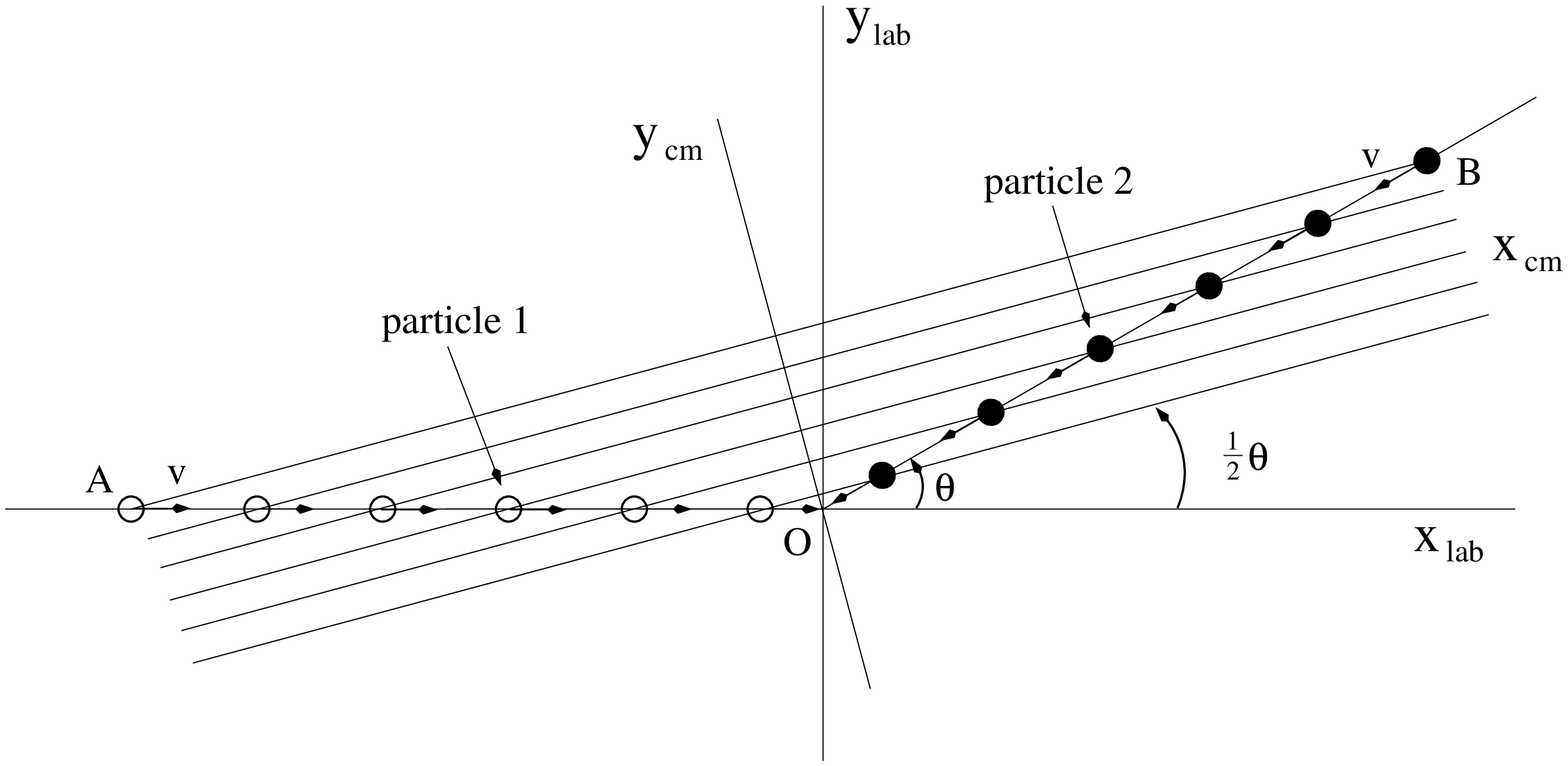}
\end{center}
\caption{This figure shows the positions of the two particles as they move toward one another before the collision.}
\label{fig_04-27-16:lab_and_cm_frames}
\end{figure}
This figure shows the tracks of the particles in the ``lab'' frame whose origin is chosen to be at the point where the particles meet and the $x_{lab}$ axis is chosen along the direction of particle 1.  The track of particle 2 runs through the origin and makes an angle $\theta$ with the $x_{lab}$ axis.  In the figure the particle positions are shown at six different times before the collision.  The pair of particle locations corresponding to a given time are connected by a line.  These lines denote the $x_{cm}$ axis of the CM frame of reference.  The line perpendicular to these lines is the $y_{cm}$ axis in this frame.  

The $x_{cm}$ axis is inclined from the $x_{lab}$ axis by angle $\theta/2$.  This can be seen by considering the triangle AOB in the figure which runs from the open--circle particle furthest from the origin (point A) to the filled--circle particle furthest from the origin (point B) to the origin (point O) and back to point A.  Since the particles are equidistant to the origin, this is an isosceles triangle.  Hence angles BAO and ABO are equal and the sum of these angles is supplemental to angle AOB.  But, since angle AOB is also supplemental to $\theta$ it follows that the sum of the two equal angles BAO and ABO is equal to $\theta$ so both angles BAO and ABO equal $\theta/2$. 
  
Thus we can define unit vectors for the lab $\hat{\mathbf i}_{lab}$ and $\hat{\mathbf j}_{lab}$ that point along the $x_{lab}$ and $y_{lab}$ axes respectively.  Additionally we can define unit vectors that point along the $x_{cm}$ and $y_{cm}$ and denote them as $\hat{\mathbf i}_{cm}$ and $\hat{\mathbf j}_{cm}$, respectively.  Although the CM frame is moving with respect to the lab frame, the relative orientation of the two pairs of axes remains fixed with the CM axes rotated with respect to the lab axes by the angle $\theta/2$.  Thus the CM unit vectors can be expressed in terms of the lab unit vectors as follows:
\begin{eqnarray}
\hat{\mathbf i}_{cm} 
&=&
\hat{\mathbf i}_{lab}\cos(\theta/2) + 
\hat{\mathbf j}_{lab}\sin(\theta/2)\nonumber\\
\hat{\mathbf j}_{cm}
&=&
-\hat{\mathbf i}_{lab}\sin(\theta/2) +
\hat{\mathbf j}_{lab}\cos(\theta/2).
\end{eqnarray}
With these two sets of unit vectors, we can now analyze the collision.

We will analyze the collision by requiring that the total momentum and total kinetic energy be conserved.  Our strategy will be to conserve these quantities in the CM frame.  Thus we will write down the initial velocities of particles 1 and 2 in the lab frame, transform these to the CM frame, find the final velocities there, transform the velocities back to the lab frame, and then compute the final kinetic energies in that frame.  In this way we will be able to calculate the difference in the final kinetic energies to learn how much kinetic energy can be transferred due to the collision.  Before we implement this procedure, we need to know how to transform between the two frames.

Consider two particles of masses $m_{1}$ and $m_{2}$ with position vectors ${\mathbf r}_{1}^{(lab)}$ and ${\mathbf r}_{2}^{(lab)}$.  In what follows the superscript in a variable name will denote the reference frame to which the quantity is referred.  Thus, ${\mathbf r}_{1}^{(lab)}$ is the vector that stretches from the origin of the lab frame to the location of mass $m_{1}$ while ${\mathbf r}_{1}^{(cm)}$ would be the vector that stretches from the origin of the CM frame to $m_{1}$. The position and velocity of the center of mass of the two--particle system, referenced to the lab frame, is given by
\begin{eqnarray}
{\mathbf r}_{cm}^{(lab)} 
&=& 
\frac{1}{2}
{\mathbf r}_{1}^{(lab)} +
\frac{1}{2}
{\mathbf r}_{2}^{(lab)}\nonumber\\
{\mathbf v}_{cm}^{(lab)} 
&=&
\frac{1}{2}
{\mathbf v}_{1}^{(lab)} +
\frac{1}{2}
{\mathbf v}_{2}^{(lab)}
\label{rcm_vcm}
\end{eqnarray}
where we have assumed $m_{1}=m_{2}$. This gives the velocity of the CM frame as measured by an observer in the lab frame.  We can transform velocities between the lab and CM frames using
\begin{eqnarray}
{\mathbf v}^{(lab)} &=& 
{\mathbf v}_{cm}^{(lab)} +
{\mathbf v}^{(cm)}.
\end{eqnarray}
We will use this equation to express the initial velocities of the colliding particle with respect to the CM frame.

From the picture of the collision in Fig.\ \ref{colliding_atoms} we can write the velocities of the colliding particles in the lab frame.  These initial velocities are
\begin{eqnarray}
\mathbf{v}_{1i}^{(lab)} 
&=& 
v\,\hat{\mathbf{i}}_{lab}\nonumber\\
\mathbf{v}_{2i}^{(lab)} 
&=& 
-v\cos(\theta)\,\hat{\mathbf{i}}_{lab} -
v\sin(\theta)\,\hat{\mathbf{j}}_{lab}.
\end{eqnarray}
These velocities can now be used to compute the velocity of the center of mass in the lab frame:
\begin{eqnarray}
\mathbf{v}_{cm}^{(lab)} = 
\frac{1}{2}\mathbf{v}_{1i}^{(lab)} + \frac{1}{2}\mathbf{v}_{2i}^{(lab)} 
&=& 
\frac{1}{2}v\left(1-\cos(\theta)\right)\,\hat{\mathbf{i}}_{lab} -
\frac{1}{2}v\sin(\theta)\,\hat{\mathbf{j}}_{lab}\nonumber\\
&=&
\left(-v\sin(\tfrac{1}{2}\theta)\right)\,\hat{\mathbf{j}}_{cm}.
\end{eqnarray}
Now we can express the initial velocities with respect to the CM frame.

The initial velocities of particles 1 and 2 relative to the CM frame can be calculated as follows: 
\begin{eqnarray}
\mathbf{v}_{1i}^{(cm)} 
&=& 
\mathbf{v}_{1i}^{(lab)} - \mathbf{v}_{cm}^{(lab)}\nonumber\\
&=&
v\,\hat{\mathbf{i}}_{lab} -
\left(
\frac{1}{2}v\left(1-\cos(\theta)\right)\,\hat{\mathbf{i}}_{lab} -
\frac{1}{2}v\sin(\theta)\,\hat{\mathbf{j}}_{lab}
\right)\nonumber\\
&=&
v\cos(\tfrac{1}{2}\theta)\,\hat{\mathbf{i}}_{cm},\nonumber\\
\mathbf{v}_{2i}^{(cm)} 
&=& 
\mathbf{v}_{2i}^{(lab)} - \mathbf{v}_{cm}^{(lab)}\nonumber\\
&=&
-v\cos(\theta)\,\hat{\mathbf{i}}_{lab} -
v\sin(\theta)\,\hat{\mathbf{j}}_{lab}\nonumber\\
&-&
\left(
\frac{1}{2}v\left(1-\cos(\theta)\right)\,\hat{\mathbf{i}}_{lab} -
\frac{1}{2}v\sin(\theta)\,\hat{\mathbf{j}}_{lab}
\right)\nonumber\\
&=&
-v\cos(\tfrac{1}{2}\theta)\,\hat{\mathbf{i}}_{cm}.
\end{eqnarray}
In the CM frame, the colliding particles have equal and opposite velocities along the $x_{cm}$ axis as expected.

\begin{figure}[H] 
\begin{center}
\includegraphics[width=0.80\linewidth]{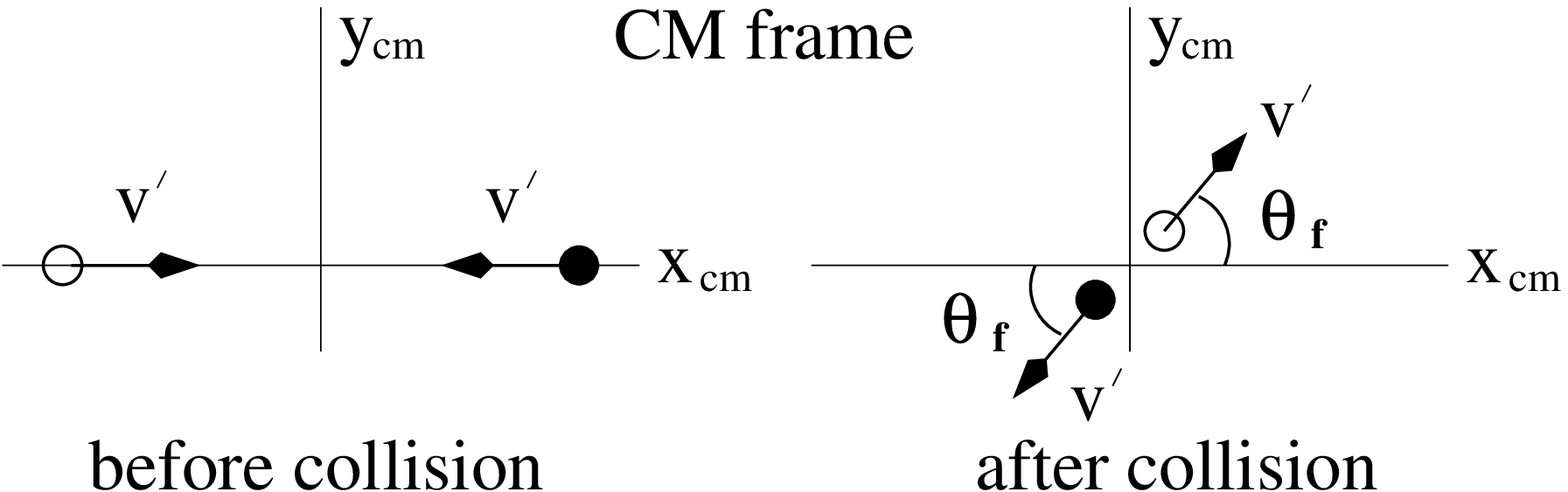}
\end{center}
\caption{Collision as observed in the CM frame.  The particle speeds before and after the collision are the same because the collision is perfectly elastic.}
\label{cm_collision}
\end{figure}

In the CM frame the total momentum of the system is zero before and after collision.  Also, since kinetic energy and total momentum are conserved in the lab frame and because the CM frame moves at a constant velocity relative to the lab frame, the total momentum and kinetic energy, as measured in the CM frame, will also be conserved.  Figure \ref{cm_collision} shows before and after pictures of the collision as observed in the CM frame.
The particles have oppositely directly velocities before and after the collision with equal speeds guaranteeing that the total momentum of the system is always zero.  The directions of their final velocities don't have to be the same as the initial velocities.  In Fig.\ \ref{cm_collision} the angle between the velocities before and after the collision is denoted as $\theta_{f}$ which we will refer to as the ``CM scattering angle''.  This angle can vary between $0^{\circ}$ and $180^{\circ}$.  The CM--frame speeds of the particles before the collision are $v^{\prime}\equiv v\cos(\tfrac{1}{2}\theta)$.  If the speed of the particles after the collision is denoted as $v^{\prime\prime}$, then conservation of kinetic energy requires that 
\begin{equation}
\frac{1}{2}M(v^{\prime})^{2} + 
\frac{1}{2}M(v^{\prime})^{2} =
\frac{1}{2}M(v^{\prime\prime})^{2} + 
\frac{1}{2}M(v^{\prime\prime})^{2}
\end{equation}
Thus we have $v^{\prime\prime} = v^{\prime} = v\cos(\tfrac{1}{2}\theta)$.

From the figure it is easy to write down the CM--frame final velocities for the two particles.  These are
\begin{eqnarray}
\mathbf{v}_{1f}^{(cm)} 
&=& 
v^{\prime}\cos(\theta_{f})\hat{\mathbf{i}}_{cm} +
v^{\prime}\sin(\theta_{f})\hat{\mathbf{j}}_{cm}\nonumber\\
\mathbf{v}_{2f}^{(cm)} 
&=& 
-v^{\prime}\cos(\theta_{f})\hat{\mathbf{i}}_{cm} -
v^{\prime}\sin(\theta_{f})\hat{\mathbf{j}}_{cm}
\end{eqnarray}
In expressing these velocities back in the lab frame and it will be convenient to express them in terms of the CM--frame unit vectors:
\begin{eqnarray}
\mathbf{v}_{1f}^{(lab)} 
&=&
\mathbf{v}_{cm}^{(lab)} +
\mathbf{v}_{1f}^{(cm)}\nonumber\\
&=&
v\cos(\tfrac{1}{2}\theta)\cos(\theta_{f})\,\hat{\mathbf{i}}_{cm} +
\left(
v\cos(\tfrac{1}{2}\theta)\sin(\theta_{f}) -
v\sin(\tfrac{1}{2}\theta)
\right)\,\hat{\mathbf{j}}_{cm}\nonumber\\
\mathbf{v}_{2f}^{(lab)} 
&=&
\mathbf{v}_{cm}^{(lab)} +
\mathbf{v}_{2f}^{(cm)}\nonumber\\
&=&
-v\cos(\tfrac{1}{2}\theta)\cos(\theta_{f})\,\hat{\mathbf{i}}_{cm} -
\left(
v\cos(\tfrac{1}{2}\theta)\sin(\theta_{f}) +
v\sin(\tfrac{1}{2}\theta)
\right)\,\hat{\mathbf{j}}_{cm}
\end{eqnarray}
The last step will be to compute the difference in lab--frame final kinetic energies.

The difference in the lab--frame kinetic energies of the two particles after the collision can be written as
\begin{equation}
\delta K^{(lab)} \equiv
\frac{1}{2}M\left(v_{2f}^{(lab)}\right)^{2} -
\frac{1}{2}M\left(v_{1f}^{(lab)}\right)^{2}.
\end{equation}
We can calculate the squares of the vectors $\mathbf{v}_{1f}^{(lab)}$ and $\mathbf{v}_{12f}^{(lab)}$ using their components in the CM--frame coordinate system since vector lengths are independent of the coordinate system.  Thus we have
\begin{eqnarray}
(v_{1f}^{(lab)})^{2} 
&=& 
v^{2} - v^{2}\sin(\theta)\sin(\theta_{f})\nonumber\\
(v_{2f}^{(lab)})^{2} 
&=& 
v^{2} + v^{2}\sin(\theta)\sin(\theta_{f})
\end{eqnarray}
Substituting these squared velocities into the expression for the kinetic--energy difference the result is
\begin{eqnarray}
\delta K^{(lab)} 
&=&
\frac{1}{2}M\left(v^{2} + v^{2}\sin(\theta)\sin(\theta_{f})\right) -
\frac{1}{2}M\left(v^{2} - v^{2}\sin(\theta)\sin(\theta_{f})\right)\nonumber\\
&=&
\frac{1}{2}Mv^{2}
\left(
2\sin(\theta)\sin(\theta_{f})
\right).
\end{eqnarray}
The final kinetic energies in the lab frame can be written as
\begin{equation}
K_{1f}^{(lab)} = \frac{1}{2}Mv^{2}
\left(
1 - \sin(\theta)\sin(\theta_{f})
\right)
\end{equation}
and
\begin{equation}
K_{2f}^{(lab)} = \frac{1}{2}Mv^{2}
\left(
1 + \sin(\theta)\sin(\theta_{f})
\right).
\end{equation}
We can now find the maximum possible transfer of kinetic energy in a collision of the type considered here.  Since $\theta_{f}$ can take any value from $0^{\circ}$ to $180^{\circ}$, the maximum and minimum kinetic energies for fixed $\theta$ occur when $\theta_{f}=90^{\circ}$:
\begin{eqnarray}
K_{1f,min}^{(lab)} 
&=& 
\frac{1}{2}Mv^{2}\left(1 - \sin(\theta)\right) =
K_{1i}^{(lab)} \left(1 - \sin(\theta)\right)\nonumber\\
K_{2f,max}^{(lab)} 
&=& 
\frac{1}{2}Mv^{2}\left(1 + \sin(\theta)\right) =
K_{2i}^{(lab)}\left(1 + \sin(\theta)\right)
\label{final_energies}
\end{eqnarray}
We will use these expressions to determine if a single collision between wall--bounce and straight--through atoms can transfer enough kinetic energy to enable atoms to escape the drain well.

\end{document}